\documentclass[a4paper, conference]{IEEEtran}
\usepackage{tabularx}
\usepackage{lipsum} 
\usepackage[utf8]{inputenc} 

\usepackage{algorithm}
\usepackage[noend]{algorithmic}
\usepackage{amsfonts}
\usepackage{amsmath}
\usepackage{amssymb}
\usepackage{amsthm}
\usepackage{array}
\usepackage{bbm, dsfont}
\usepackage{boxedminipage}
\usepackage{caption}
\usepackage{color}
\usepackage{colortbl}
\usepackage{cuted}
\usepackage{diagbox}
\usepackage{enumitem}
\usepackage{fancyhdr}
\usepackage{floatflt}
\usepackage{float}
\usepackage{graphics}
\usepackage{graphicx}
\usepackage{indentfirst}
\usepackage{latexsym}
\usepackage{lineno}
\usepackage{listings}
\usepackage{mathtools}
\usepackage{multirow}
\usepackage{multicol}
\usepackage[square,numbers,sort,compress]{natbib}
\usepackage{notoccite}
\usepackage{picinpar}
\usepackage{placeins}
\usepackage{psfrag}
\usepackage{physics}
\usepackage{romannum}
\usepackage{soul}
\usepackage{subfigure}
\usepackage{supertabular}
\usepackage{spverbatim}
\usepackage{textcomp}
\usepackage{theorem}
\usepackage[para]{threeparttable} 
\usepackage{pdfpages}
\usepackage{theorem} 
\usepackage{tikz}
\usetikzlibrary{graphs}
\usepackage{url}
\usepackage{xcolor}

\usepackage{wrapfig}

\RequirePackage{xifthen, xparse, amsmath, amssymb, color}

\def\argmin{\mathop{\operator@font argmin}}
\def\Argmin{\mathop{\operator@font Argmin}}
\def\argmax{\mathop{\operator@font argmax}}
\def\Argmax{\mathop{\operator@font Argmax}}

\newcommand\itemeqn[1][]
{
	\ifthenelse{\isempty{#1}}
		{
		\item\abovedisplayskip=2pt\abovedisplayshortskip=0pt~\vspace*{-\baselineskip}
		}
		{
		\item[#1]\abovedisplayskip=2pt\abovedisplayshortskip=0pt~\vspace*{-\baselineskip}
		}
}

\DeclareDocumentCommand{\tn}{O{t}}{\ensuremath{_{#1+1}}}
\DeclareDocumentCommand{\tp}{O{t}}{\ensuremath{_{#1-1}}}
\DeclareDocumentCommand{\ag}{O{}O{t}}{
	\ensuremath{
		\ifthenelse{\isempty{#1}}
		{
			_{#2}^{ag}
		}
		{
			_{#2+#1}^{ag}
		}
	}
}
\DeclareDocumentCommand{\md}{O{}O{t}}{
	\ensuremath{
		\ifthenelse{\isempty{#1}}
		{
			_{#2}^{md}
		}
		{
			_{#2+#1}^{md}
		}
	}
}

\usepackage{amsthm}

\newcommand{\ignore}[1]{ }

\newcommand{\figlbl}[1]{\label{fig.{#1}}}
\newcommand{\figref}[1]{Fig.~\ref{fig.{#1}}}
\newcommand{\tbllbl}[1]{\label{table:{#1}}}
\newcommand{\tblref}[1]{Table~\ref{table:{#1}}}
\newcommand{\seclbl}[1]{\label{sec:{#1}}}
\newcommand{\secref}[1]{Section~\ref{sec:{#1}}}

\newcommand{\rmv}[1]{}

\definecolor{mygray}{gray}{0.8}

\setcitestyle{citesep={,}}
 \usepackage{amsthm}
\title{Integral Sampler and Polynomial Multiplication Architecture for Lattice-based Cryptography
}
\author{Antian Wang$^\star$, Weihang Tan$^\star$, Keshab K. Parhi$^\ddag$, Yingjie Lao$^\star$\\
$^\star$Department of Electrical and Computer Engineering, Clemson University, Clemson, SC 29634, USA\\
$^\ddag$Department of Electrical and Computer Engineering, University of Minnesota, Minneapolis, MN 55455, USA\\
$^\star$\{antianw, wtan, ylao\}@clemson.edu,
$^\ddag$parhi@umn.edu}

\usepackage{natbib}
\usepackage{graphicx}
\IEEEoverridecommandlockouts
\IEEEpubid{\makebox[\columnwidth]{978-1-6654-5938-9/22/\$31.00 \copyright2022 IEEE \hfill} \hspace{\columnsep}\makebox[\columnwidth]{ }}
\begin{document}

\maketitle

\begin{abstract}
With the surge of the powerful quantum computer, lattice-based cryptography proliferated the latest cryptography hardware implementation due to its resistance against quantum computers. Among the computational blocks of lattice-based cryptography, the random errors produced by the sampler play a key role in ensuring the security of these schemes. This paper proposes an integral architecture for the sampler, which can reduce the overall resource consumption by reusing the multipliers and adders within the modular polynomial computation. For instance, our experimental results show that the proposed design can effectively reduce the discrete Ziggurat sampling method in DSP usage. 
\end{abstract}

\maketitle
\section{Introduction}
Cybersecurity is being shaped by the need to secure algorithms, data, devices, and networks under future threats as technology evolves~\cite{lao2014protecting,clements2018backdoor}. The present in-use public-key cryptosystems, such as RSA and elliptic curve cryptography (ECC), are vulnerable to the attack of the quantum computer using Shor's algorithm~\cite{shor1999polynomial}. Currently, the National Institute of Standards and Technology (NIST) is standardizing one or more quantum-resistant public-key cryptographic algorithms, which is known as post-quantum cryptography (PQC)~\cite{bernstein2017post}. These PQC schemes can be classified as key encapsulation mechanism (KEM) based~\cite{bos2018crystals,alkim2016post} and digital signature based schemes~\cite{ducas2018crystals,banerjee2019sapphire}. Several schemes relying on the lattice-based computational primitive are selected for the round 4 standardization.

Meanwhile, the hardware implementations and accelerations for the finalist candidates are required for evaluating the overall performance. In general, the accelerations of PQC schemes focus on modular polynomial multiplier, modular (integer) multiplier, hash module, and sampler~\cite{zhu2021lwrpro,banerjee2019sapphire}. In particular, the efficient designs for modular polynomial multiplier have been extensively studied, using number theoretic transform (NTT)~\cite{tan2021pipelined,oder2017implementing,kuo2017high}, schoolbook polynomial multiplication algorithm~\cite{roy2020high,xie2021crop}, or Karatsuba algorithm~\cite{mera2020compact,zhu2021lwrpro}. The architectures for modular multiplier and hash module have also been investigated in~\cite{paludo2022NTT,tan2021high,sundal2017efficient,tan2021low}. In contrast, hardware architectures for the sampler are less studied. The PQC schemes deploying the learning with errors (LWE) problem's variants require adding the errors into the ciphertext to make the problem computationally hard. Such errors are typically generated from the sampler, which protects secure information.

Different from previous architectures that optimize the sampler architecture independently~\cite{howe2018standard,sinha2013high,karmakar2018constant,lyons2020sampling,zhang2005ziggurat,agrawal2020post}, this paper proposes a novel integral architecture by optimizing the sampler jointly with the modular polynomial multiplier, accompanied by a simple control unit. The sampler in the proposed design utilizes the abundant computation resources within the underlying modular polynomial multiplier to reduce the overall resource consumption. In particular, we design the Gaussian distribution samplers using the Knuth-Yao algorithm~\cite{knuth1976complexity} and discrete Ziggurat algorithm~\cite{buchmann2013discrete}, respectively, by following the proposed integral architecture. 

The rest of this paper is organized as follows: \secref{back} reviews the mathematical background for lattice-based cryptography and the corresponding hardware architectures in prior works. \secref{design} presents the details of our hardware architecture design. The performance of our proposed architecture is provided and analyzed in \secref{exp}. Finally, \secref{conclu} presents remarks and concludes the paper.

\begin{figure*}[t]
    \centering
    \resizebox{0.7\textwidth}{!}{
    \includegraphics{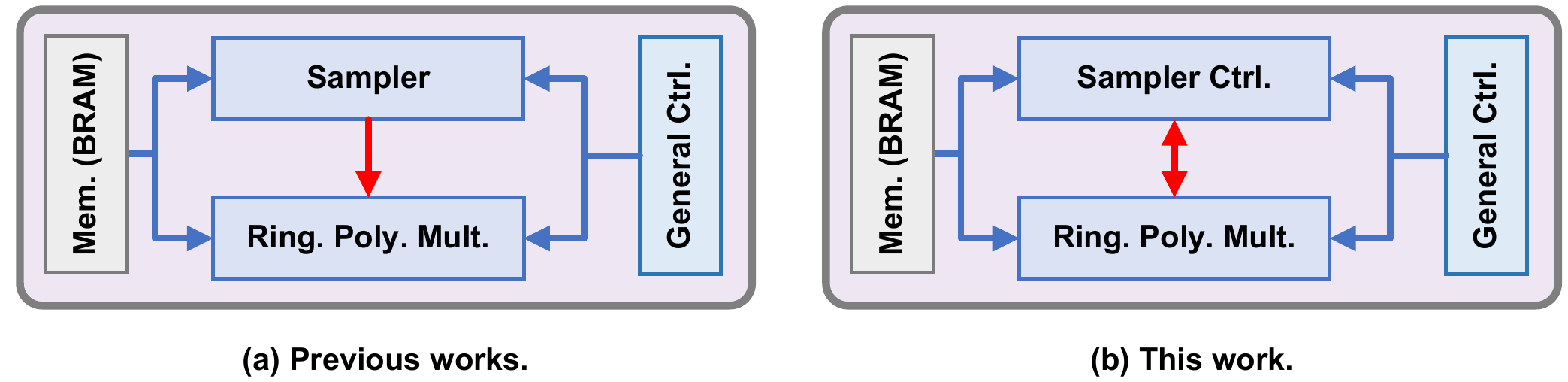}}
    \caption{Comparison with prior works.}
    \figlbl{scheme_comp}
    \vspace{-1em}
\end{figure*}

\section{Background} \seclbl{back}
LWE is based on the lattice problem, which is NP-hard for quantum computers~\cite{regev2009lattices}. Due to its arithmetic simplicity and strong security performance, PQC schemes based on the LWE problem's successors are popular among the existing schemes. Crystals-Kyber (Module-LWE)~\cite{bos2018crystals} and NewHope (Ring-LWE)~\cite{alkim2016post}, are two representative LWE-based cryptosystems that have been considered for the NIST PQC standardization. 

For a Ring-LWE based scheme, the polynomial computation is computed over ring $R_q = \mathbb{Z}_{q}/(x^n+1)$, where $x^n+1$ is an irreducible polynomial of degree (dimension) $n$, and the polynomial coefficient modulus is $q$. The Ring-LWE sample $(a(x),b(x)) \in R_q \times R_q$ is defined as follows: $a(x)$ is an uniformly random polynomial over ring $R_q$, and the corresponding $b(x)$ is expressed as 
\begin{equation}
    b(x) = a(x)\cdot s(x) +e(x) \in R_q,
\end{equation}
where $s(x) \in R_q$ is the secret, and $e(x) \in R_q$ is the error term. Specifically, both polynomials are random polynomials sampled from a discrete Gaussian distribution with a standard deviation of $\sigma$~\cite{lyubashevsky2010ideal}. Notably, the construction of Ring-LWE and Module-LWE based variants are similar with the only difference of $s(x)$ and $a(x)$ are $d$-dimension vectors, i.e., $s(x), a(x) \in (R_q)^d$, and the entries of $s(x)$ and $a(x)$ are polynomials in $R_q$.

Meanwhile, the multiplication between two polynomials (vectors) over the ring is heavily used in LWE-based schemes. In the literature, the number-theoretic transform (NTT) based modular polynomial multiplication is widely used to reduce the quadratic complexity $\mathcal{O}(n^{2})$ from the schoolbook polynomial multiplication to $\mathcal{O}(n\log n)$~\cite{tan2021pipelined}. The main procedure of the modular polynomial multiplication of two polynomials using NTT  is to convert all the coefficients of the polynomials into the NTT domain. Then a direct coefficient-wise multiplication is performed between the two polynomials. Finally, the resulting polynomial uses an inverse NTT to recover the coefficients back to the original algebraic domain polynomial.

\section{Proposed Sampler Designs}\seclbl{design}
In this section, we present the hardware architecture designs for two widely used sampling algorithms, i.e., the Knuth-Yao algorithm and the discrete Ziggurat algorithm.

\subsection{High-level overview}\seclbl{highlevel}

Our design aims to reduce the hardware cost by sharing the arithmetic operations between the sampling and ring modular polynomial multiplication (i.e., NTT) to facilitate an integral architecture. Besides, since ring modular polynomial computations need the results from the sampler for key generation and encryption, sharing the hardware resources does not impact the overall computation flow. The difference between this work and prior works is shown in~\figref{scheme_comp}. Previous works typically separate sampler from the ring modular polynomial multiplier as depicted in~\figref{scheme_comp}(a). In contrast, we design the sampler as an integral part of the modular polynomial architecture as shown in~\figref{scheme_comp}(b). Here the sampler is redesigned as a sampling control module, where the sampling process relies on the results computed by the ring modular polynomial multiplier rather than computed within the sampler module. Hence, it can reduce the resource consumption of sampling algorithms involving multiplications and additions. 

The sampling control module is responsible for providing the ring modular polynomial multiplication with the intermediate value during sampling computation and sampled value for key generation and encryption. The detailed reconfiguration process for a typical ring modular polynomial multiplication using NTT is shown in~\figref{butconfig}. The reconfiguration uses multiplexers (MUXes) to configure the NTT unit for ordinary NTT operation or reconfigured sampling operation. For sampling operation, the paths of the butterfly unit are only partially activated for addition or multiplication. The modular addition consists of a general addition and modular reduction, which is used to reduce the results to be smaller than $q$. Modular addition can be configured to general addition by bypassing modular reduction. General additions with longer bit-length than $q$ can be realized by cascading several general adders reconfigured from modular adders. For modular multiplications, it first multiplies two products and then reduces the final result to be less than $q$. Similarly, configuring modular multiplication to general multiplication can be realized by extracting the multiplication result without modular reductions. Given that the sampling algorithm could have large bit-length multiplications that cannot be accommodated within one modular polynomial multiplier unit (i.e., NTT unit), we need to partition them into small bit-length multiplications using general multipliers reconfigured from modular multipliers. Then, we will add the partial results based on their actual locations within the large bit-length multiplication results to obtain the final product using general adders reconfigured from modular adders. 

\begin{figure*}[htbp]
    \centering
     \resizebox{0.8\textwidth}{!}{
    \includegraphics{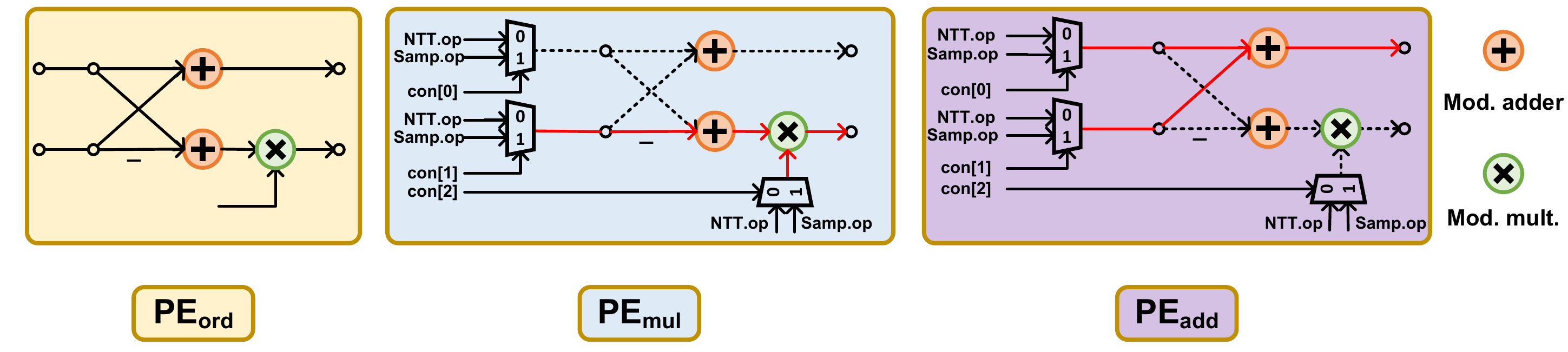}}
    \caption{Butterfly unit configuration. From left to right: general butterfly unit, butterfly unit reconfigured to multiplication, butterfly unit reconfigured to addition. Red lines and dashed lines show the activated and inactivated data paths, respectively.}
    \figlbl{butconfig}
\end{figure*} 

\subsection{Knuth-Yao sampling control design}
\subsubsection{Sampling algorithm background}
A tree-based sampling algorithm for non-uniform distributions was proposed in~\cite{knuth1976complexity}. It builds a probability matrix $P$ for non-negative values' probability for $0$ to $N-1$ as $p_0,\hdots p_{N-1}$  represented with a maximum $\lambda$ bits individually. It can also be represented as a discrete distribution tree with maximum $\lambda$ levels with internal nodes that have two children and terminal nodes. The sampling process starts from the root of the tree to the terminal nodes by iteratively consuming a single random bit and stepping down one level until it reaches a terminal node. The integer label of the reached terminal node will be returned as the sampled value. The process is described in Algorithm~\ref{KYMethod}. 

\begin{algorithm}
\caption{Knuth-Yao Algorithm~\cite{howe2016practical}}
\begin{algorithmic}[1]\label{KYMethod}
\REQUIRE Probability matrix $P$ with size $N\times \lambda$. Column-wise Hamming distance for $P$ as $HD[j]=\sum\limits_{i=0}^{N-1}P[i][j]$. Random bit array $r$.
\ENSURE Sample value $(-1)^{r[cnt]}\cdot row$.
\STATE $d=0,cnt=0$
\FOR{$col=0,col<\lambda,col=col+1$}
\STATE $d=2d+(!r[cnt])-HD[col]$
\STATE $cnt=cnt+1$
\IF{$d<0$}
\FOR{$row=0,row<N,row=row+1$}
\STATE $d=d+P[row][col]$
\IF{$d=0$}
\STATE return $(-1)^{r[cnt]}\cdot row$
\ENDIF
\ENDFOR{}
\ENDIF
\ENDFOR{}
\end{algorithmic}
\end{algorithm}

Various hardware implementations and optimizations for the Knuth-Yao algorithm have been studied. For instance, a hardware-friendly random walk for discrete Gaussian sampler using the Kunth-Yao algorithm was implemented in~\cite{sinha2013high}. A reconfigurable adder tree for the Hamming distance computation and loop unrolling for the sample search was proposed in~\cite{zhao2022discrete}. A discrete Gaussian sampler exploits the $i$-th column/tree-level information to identify the nodes for tree traversal instead of visiting each level at the random walk was proposed in~\cite{roy2014compact}. 
The observation based on the unique mapping between the input random bits and the output samples allows each output sample bit to be produced by a Boolean function of the input bit sequence, inducing the constant-time Gaussian sampling proposed in~\cite{karmakar2018constant}. A custom combinational circuit generation for sampling from arbitrary complete discrete distributions was proposed in~\cite{lyons2020sampling}. However, all these prior works design the sampler separately from the polynomial multiplier.

\subsubsection{Modular polynomial multiplier configuration}
The main computations of the Knuth-Yao algorithm are at line 3 and line 6 of Algorithm~\ref{KYMethod}. Given $d=2d+(!r[cnt])-HD[col]$, $d=d+P[row][col]$, where $0\leq HD[col]\leq \lambda, P[row][col]\in \{0,1\},0\leq d<2^\lambda$, we can configure modular addition to accommodate the update procedure of $d$ by cascading several general adders. Additionally, as seen from the operation at line 3 in Algorithm~\ref{KYMethod}, $d$ should always be less  than $HD_{sum}=\sum\limits_{col=0}^{\lambda-1} HD[col]$ to have a valid sample value. 
It can further reduce the required total bit-length of general adder configured from modular adder to support sampling operations from $\lambda+1$ bits to $(\log HD_{sum})+1$ bits. Even without the optimization, for example, for distribution with $\sigma=215$ and $q=11289$ with the maximum sample value of $9\sigma=1935$, the maximum possible $HD_{sum}$ is $1935\times 64=12840$, which can be represented in $14$ bits. 

\subsubsection{Sampler control design} 
The overall hardware design for the Knuth-Yao algorithm is shown in~\figref{KYmethod}. The control module that determines the sampling process consists of three components, i.e., $d<0$ at line 4 by checking the most significant bit of $d$ that is represented in $2$'s complement, $d=0$ at line 7 in Algorithm~\ref{KYMethod}, and  $d>HD_{sum}$ to allow early stopping as discussed above. The value of $d$ is updated based on these three comparisons to determine the following update pattern, or output sampled value $(-1)^{r[cnt]}\cdot row$ when $d=0$.

\begin{figure}[ht]
    \centering
        \resizebox{0.35\textwidth}{!}{
    \includegraphics{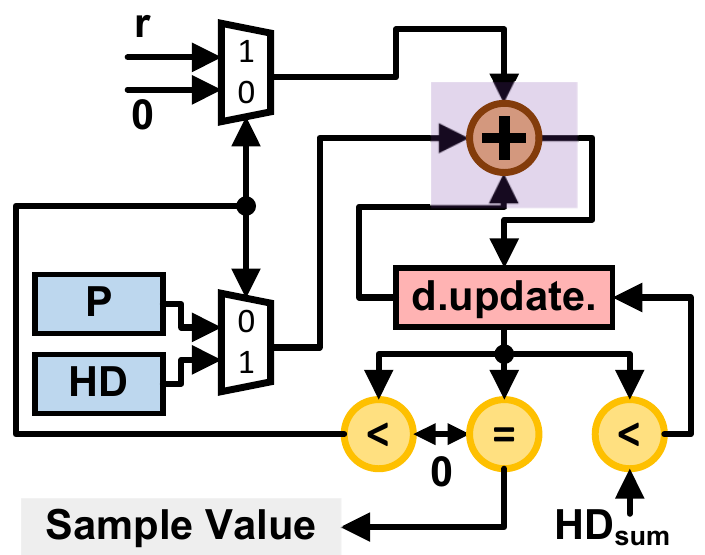}}
    \caption{Knuth-Yao sampler hardware design. The shaded adder is computed within the ring modular polynomial module.}
    \figlbl{KYmethod}
\end{figure}

\subsection{Discrete Ziggurat sampling control design}
\subsubsection{Sampling algorithm background}
Rejection sampling is another category of sampling algorithm applicable for an arbitrary probability distribution. 
Ziggurat sampling is an optimized form of rejection sampling by covering the probability density function with several horizontal rectangles with identical areas~\cite{marsaglia2000Ziggurat}. 
For the selected rectangle with index $i$, points to the left of $x_{i-1}$ are accepted, while each sampled point to the right need further computation based on its location over the probability distribution function to determine whether it gets accepted or not. The later computations generally incur a large overhead. The discrete Ziggurat~\cite{buchmann2013discrete} algorithm is presented in Algorithm~\ref{disZiggurat} with \textit{sLine} operation described in Algorithm~\ref{disZigguratsline}. 

\begin{algorithm}[ht]
\caption{Discrete Ziggurat Algorithm~\cite{buchmann2013discrete}}
\begin{algorithmic}[1]\label{disZiggurat}
\REQUIRE The number of rectangles $m$. Lower right corners for rectangles with coordinates $\lfloor x_i\rfloor,\bar y_i=\lfloor y_i\rfloor$ for $i\in m$ with precision $\lambda$.
\ENSURE $s\cdot x$.
\WHILE{True}
\STATE choose rectangle $i\in \{1,\hdots m\}$, sign $s\in \{-1,1\}$, random bit $b\in \{0,1\}$, choose random value $x\in \{0,\hdots, \lfloor x_i\rfloor \}$ and $y'\in \{0,\hdots 2^\lambda -1\}$
\IF{$0<x\leq \lfloor x_{i-1}\rfloor$}
\STATE return $s\cdot x$
\ELSE 
\IF{$x=0$ \AND $b=0$}
\STATE return $s\cdot x$
\ELSE
\STATE $\bar y=y'(\bar y_{i-1}-\bar y_i)$
\STATE $y_{s}= \textit{sLine}(\lfloor x_{i-1}\rfloor$,$\lfloor x_{i}\rfloor$, $\bar y_{i-1}$, $\bar y_i$, $x$)
\IF{$\lfloor x_i\rfloor+1\leq \sigma$}
\IF{$\bar y\leq 2^\lambda \cdot y_{s}$ \AND $\bar y\leq 2^\lambda\cdot (p_x-\bar y_i)$}
\STATE return $s\cdot x$
\ELSE \STATE continue
\ENDIF
\ELSE
\IF{$\sigma\leq \lfloor x_{i-1}\rfloor $}
\IF{$\bar y\geq 2^\lambda\cdot y_{s}$ \AND $\bar y>2^\lambda\cdot (p_x-\bar y_i)$}
\STATE continue
\ELSE
\STATE return $s\cdot x$
\ENDIF
\ELSE
\IF{$\bar y\leq 2^\lambda(p_x-\bar y_i)$}
\STATE return $s\cdot x$.
\ELSE
\STATE continue
\ENDIF
\ENDIF
\ENDIF
\ENDIF
\ENDIF
\ENDWHILE
\end{algorithmic}
\end{algorithm}

\begin{algorithm}
\caption{\textit{sLine}() Operation in Algorithm \ref{disZiggurat}}
\begin{algorithmic}[1]\label{disZigguratsline}
\REQUIRE $\lfloor x_{i-1}\rfloor$,$\lfloor x_{i}\rfloor$, $\bar y_{i-1}$, $\bar y_i$, $x$
\IF{$\lfloor x_{i-1}\rfloor=\lfloor x_{i}\rfloor$}
\STATE return $-1$
\ENDIF
\STATE 
$\hat y_i=\bar y_i$, $\hat y_{i-1}=\begin{cases}
\bar y_{i-1},i>1\\
1,i=1
\end{cases}$
\STATE return $\frac{\hat y_i-\hat y_{i-1}}{\lfloor x_{i}\rfloor-\lfloor x_{i-1}\rfloor}(x-\lfloor x_{i}\rfloor)$
\end{algorithmic}
\end{algorithm}

One of the first hardware Ziggurat sampling implementations was proposed in~\cite{zhang2005ziggurat}. The design pipelines the computation of rectangular regions in parallel with the evaluation of the wedge and tail, which are the areas underneath the probability distribution to the right of $x_{i-1}$. A variant of Ziggurat sampling that uses trapezoids rather than rectangles to partition the distribution was proposed in~\cite{edrees2009hardware}. The partition leaves fewer rejection areas with similar hardware performance compared with the rectangle partition. 
A parameterized design for Ziggurat sampling was proposed in~\cite{agrawal2020post} explores the trade-offs between the number of rectangles for partition and hardware resource consumption. Recently, a discrete Ziggurat sampling hardware implementation is proposed~\cite{howe2016practical}. The detailed comparisons over acceptance and rejection regions are summarized in Algorithm~\ref{disZiggurat}. The sampling algorithm is the major computational bottleneck.

\begin{figure}
    \centering
 \resizebox{0.48\textwidth}{!}{
    \includegraphics{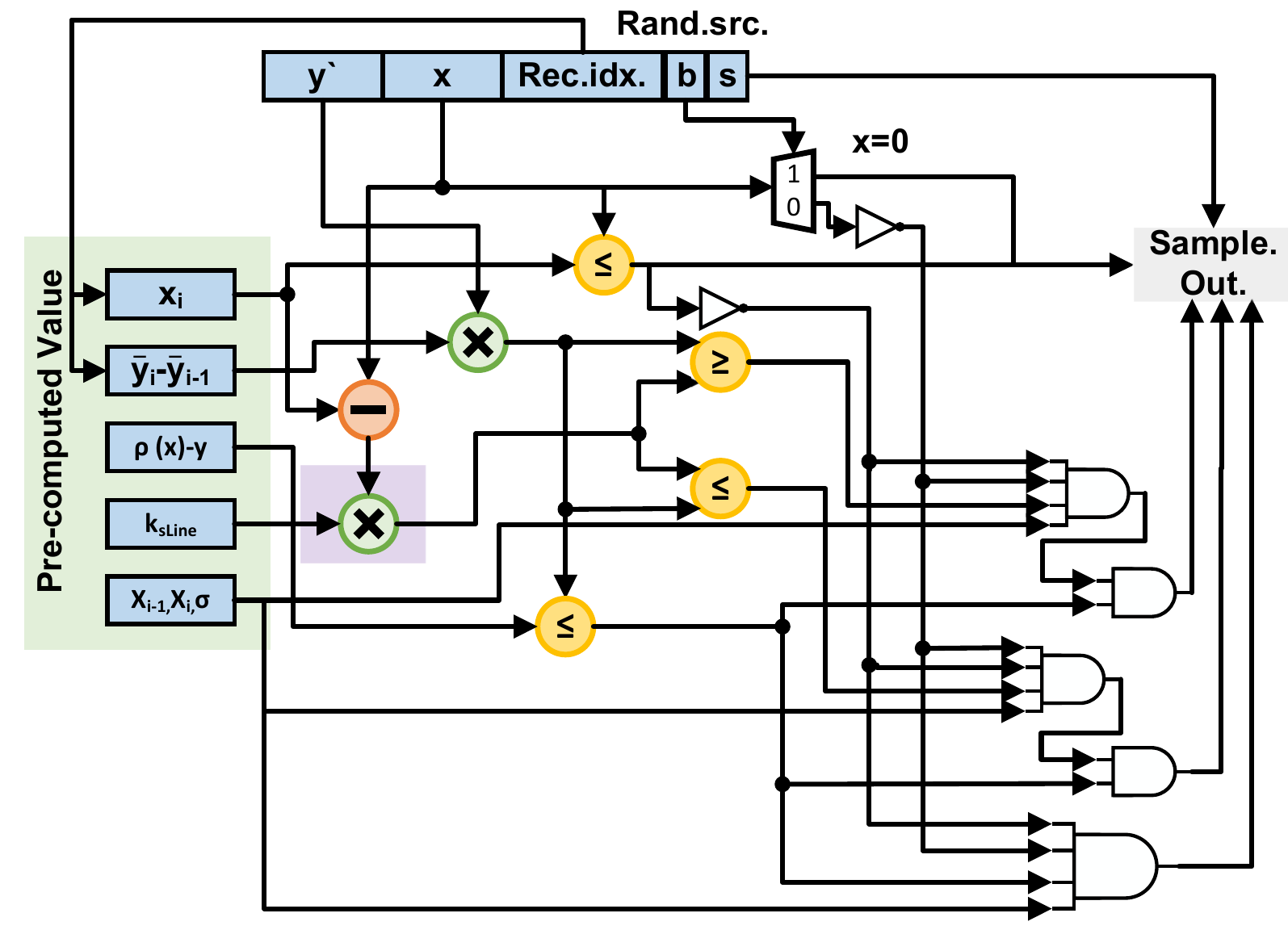}}
    \caption{Discrete Ziggurat sampler hardware design. The shaded multiplier is computed within the ring modular polynomial module.}
    \figlbl{Zigmethod}
    \vspace{-1em}
\end{figure}

\begin{table*}[htbp]
    \centering
     \caption{Performance of the Knuth-Yao Sampler (Post Place and Route)
     }
\begin{tabular}{cccccccc}
Distribution parameter and setting &Module&LUT/FF/Slice&BRAM/DSP\\
\hline
$\sigma_{LP}= 3.33$ w/o. config
&NTT& 1304/1368/1476&0/18  \\ 
&Sampler&  64/64/28 &2/0 \\
\hline
$\sigma_{BLISS}= 215$ w/o. config&NTT&3486/2109/3852&0/10   \\ 
&Sampler& 94/41/94&7/0  \\
\hline
$\sigma_{LP}= 3.33$ w. config &Reconfigured NTT &1323/1628/1491  &0/18  \\ 
&Sampler Control&71/43/71&2/0\\
\hline
$\sigma_{BLISS}= 215$ w. config &Reconfigured NTT   &
3513/2371/3865
&0/10\\
&Sampler Control&43/51/29&7/0\\
\hline
\end{tabular}
     \tbllbl{KY}
\end{table*}

\begin{table*}[htbp]
\centering
 \caption{Performance of the Discrete Ziggurat Sampler (Post Place and Route)
    }
\begin{tabular}{cccccccc}
Distribution parameter and setting &Module&LUT/FF/Slice&BRAM/DSP\\
\hline
$\sigma_{LP}= 3.33$ w/o. config
&NTT&1303/172/1475&0/18   \\ 
&Sampler& 341/341/100&0/17 \\
\hline
$\sigma_{BLISS}= 215$ w/o. config &NTT&3625/2218/3991&0/10 \\
&Sampler& 281/107/281&6/16\\
\hline
$\sigma_{LP}= 3.33$ w. config&Reconfigured NTT&1511/1616/1679&0/18   \\
&Sampler Control&211/10/211&0/1\\
\hline
$\sigma_{BLISS}= 215$ w. config&Reconfigured NTT&3756/2258/4122&0/10\\
&Sampler Control&181/19/123&6/2  \\
\hline
\end{tabular}
   \tbllbl{Zig}
\end{table*}

\subsubsection{Modular polynomial multiplier configuration}
For the discrete Ziggurat sampler, the major computation costs lie in \textit{sLine} of Algorithm~\ref{disZigguratsline}, which determines whether the sampled point is underneath the curve or not. Given that $k=\frac{\hat y_i-\hat y_{i-1}}{\lfloor x_{i}\rfloor-\lfloor x_{i-1}\rfloor}$ is fixed for $x$ located within the selected partitioned rectangle, it reduces to $k(x-\lfloor x_{i}\rfloor)$. Thus, it can be computed by reconfiguring the modular polynomial multiplier as discussed in~\secref{highlevel}. The numbers of small bit-length general multiplications and general additions are determined by the size of $q$ and the probability distribution that the sampler follows. The reconfiguration is trivial. For $\sigma=3.33, q=4093, n=512 $, the maximum possible sampled value for the sampler is $30\approx 9\sigma$. For a $64$-bit $k$, $64$-bit $\times 5$-bit  multiplication is needed for the \textit{sLine} operation. Given that $5$-bit is smaller than $\log_2(4093)=12$ bit, we partition the $64$ bits into $6$ $12$-bit partial multiplications with an output bit-length of $17$. Then we need to accommodate $5$ general additions of bit-length $12$ to obtain the final result. Each result of the $17$-bit multiplications is partitioned into higher $5$ bits and lower $12$ bits, while the higher $5$ bits are added to the next lower $12$ bits. 
When $n=512$, we have $9$ NTT units available for the modular polynomial multiplication. Besides, we have two modular multipliers before the NTT units as part of the negative wrapped convolution as discussed in~\cite{tan2021pipelined}. To accommodate the multiplication, we reconfigure the two modular multipliers for the negative wrapped convolution as the general multiplier and then reconfigure the first $4$ NTT units as general multipliers and the last $5$ NTT units as general adders. 
Other computations in Algorithm~\ref{disZiggurat} use pre-computed values for further reducing the resource consumption, similar to the approach in~\cite{howe2018standard}. 

\subsubsection{Sampler control design}
The overall hardware design for the discrete Ziggurat algorithm is shown in~\figref{Zigmethod}. Similarly, the discrete Ziggurat sampler control module involves multiple comparisons to determine whether the sampled value is accepted or not. Each acceptable condition is inferred from the comparison results based on the computations described in Algorithm~\ref{disZiggurat}. For the comparisons at line 11 and line 17, the results are determined by the selected rectangle index $i$. A stored bit value for an individual rectangle index $i$ can be used to avoid actual comparisons. Besides, the determinations of acceptance at lines 3, 6, 12, 18, and 23 in Algorithm~\ref{disZiggurat} are done by the comparisons. All Boolean comparison results are logically OR-ed to obtain the final accepted sample value. 
\section{Experimental Results}\seclbl{exp}

In our experiment, we adapt the design in~\cite{tan2021pipelined} as our modular polynomial multiplier module. The sampler controller is fed with uniformly sampled random bits. We implement the Knuth-Yao algorithm and the discrete Ziggurat algorithm using two sets of LWE parameters $\sigma_{LP}=3.33$, $q=4093$, $n=256$~\cite{lindner2011better} and $\sigma_{BLISS}=215.73$, $q=12289$, $n=512$~\cite{ducas2013lattice}. Using the proposed design method, both achieved the operating frequency at 100MHz using the Artix-7 AC701 evaluation board. Note that $q=4093$ cannot apply the optimization for $q$ as discussed in~\cite{tan2021pipelined}, resulting in $2$ DSP units for the Barrett reduction. The performances without and with configuration are presented in \tblref{KY} and \tblref{Zig}, respectively. From \tblref{KY}, we can observe the limited difference between design without reconfiguration and design with a reconfiguration for identical parameter sets. It is reasonable since the Knuth-Yao algorithm only includes short bit-width additions. The reconfiguration logic for offloading it to the reconfigured NTT module is approximately equivalent to the short bit-width addition. From \tblref{Zig}, we can see the reconfiguration successfully offloads the expensive DSP resource usage to the reconfigured polynomial modular multiplier, with limited impact over types of resource usage. The results are consistent with the Knuth-Yao algorithm's superiority over discrete Ziggurat as concluded in~\cite{howe2016practical}. 

\section{Conclusion}\seclbl{conclu}
This paper presented an integral hardware architecture for lattice-based cryptography that integrates the sampler with the underlying ring polynomial computation unit. This design implements the widely used Kunth-Yao algorithm and Ziggurat algorithms for discrete Gaussian sampling. 
\bibliographystyle{IEEEtran}
\bibliography{00_main}
\end{document}